\documentclass[prd,twocolumn,showpacs,amsmath,amssymb,superscriptaddress,floatfix,nofootinbib]{revtex4}
\usepackage{mathrsfs,bm}
\usepackage{longtable,lscape,diagbox}
\usepackage{txfonts}
\usepackage{amssymb}
\usepackage{indentfirst}
\usepackage{graphicx,,booktabs}
\usepackage{multirow}
\usepackage{color}
\usepackage{amssymb}
\definecolor{cover}{rgb}{0.77,0.87,0.88}
\definecolor{blueone}{rgb}{0.1,0.1,.7}
\definecolor{citec}{rgb}{0.14,0.47,0.09}
\definecolor{two}{rgb}{0.0,0.5,0.}
\definecolor{three}{rgb}{.5,.1,0.15}
\usepackage[bookmarks=true,bookmarksopen=false,plainpages=false,breaklinks=true,
   bookmarksnumbered=true,hypertexnames=false,
   filecolor=blue,urlcolor=three,menucolor=three,
   linkcolor=three,citecolor=blueone, colorlinks,
   anchorcolor=blue,runcolor=pink,frenchlinks=red
   pdfstartview=FitH,pdftitle=title,%
   pdfauthor=author]{hyperref}

\def\half{{\textstyle{1\over 2}}}

\def\thalf{{\textstyle{3\over 2}}}
\def\fhalf{{\textstyle{5\over 2}}}

\usepackage{relsize}
\usepackage{xspace}
\def\babar{\mbox{\slshape B\kern-0.1em{\smaller A}\kern-0.1em
    B\kern-0.1em{\smaller A\kern-0.2em R}}}

\begin{document}
\title{Nucleon resonances $N(1875)$ and $N(2100)$ as strange partners of LHCb pentaquarks}

\author{Jun He}
\email{junhe@impcas.ac.cn}
\affiliation{Department of  Physics, Nanjing Normal University,
Nanjing, Jiangsu 210097, China}
\affiliation{Institute of Modern Physics,
Chinese Academy of Sciences, Lanzhou 730000, China}
\affiliation{Research Center for Hadron and CSR Physics, Lanzhou University and Institute
of Modern Physics of CAS, Lanzhou 730000,China}

\date{\today}
\begin{abstract}
	
In this work, we investigate the possibility of interpreting two nucleon
resonances, the $N(1875)$ and the $N(2100)$, as hadronic molecular
states from the $\Sigma^*K$ and $\Sigma K^*$ interactions,
respectively. With the help of  effective Lagrangians in which 
coupling constants are determined by the SU(3) symmetry,  the
$\Sigma^*K$ and $\Sigma K^*$  interactions are described by the
vector-meson and pseudoscalar-meson exchanges. With the
one-boson-exchange potential obtained,  bound states from the
$\Sigma^*K$ and $\Sigma K^*$  interactions are searched for in a
quasipotential Bethe-Saltpeter equation approach. A bound state with
quantum number $I(J^P)=1/2(3/2^-)$ is produced from the  $\Sigma^*K$
interaction, which can be identified as the $N(1875)$ listed in PDG. It can be seen as a strange
partner of the LHCb pentaquark $P_c(4380)$ with the same quantum
numbers in the  molecular
state picture.  The  $\Sigma
K^*$  interaction also produces a bound state with quantum number
$I(J^P)=1/2(3/2^-)$, which is related to  experimentally observed
$N(2100)$ in the $\phi$ photoproduction. Our results suggest that the $N(2120)$
observed in the $K\Lambda(1520)$ photoproduction and the
$N(2100)$ observed in the $\phi$ photoproduction  have different origins. The former
is a conventional three-quark state while the latter is a $\Sigma
K^*$ molecular state, which can be seen as a strange partner of the
$P_c(4450)$ with different spin parity.  

\end{abstract}


\maketitle
\section{INTRODUCTION}

The pentaquark is an important topic of hadron physics. Its history
can be tracked back to the birth of the quark model. The $\Theta$ particle
with a mass of about 1540 MeV claimed by the LEPS Collaboration
started a worldwide rush of the pentaquark study in both experiment
and theory~\cite{Nakano:2003qx}.  More precise experiments did not
confirm the LEPS observation, which made people lose enthusiasm about
the pentaquark~\cite{Close:2005pm}.  Study of the pentaquark study has
been revived
after  recent observations of  hidden-charmed $P_c(4380)$ and
$P_c(4450)$ at LHCb~\cite{Aaij:2015tga}. Many interpretations of 
the internal structure of LHCb pentaquarks have been proposed and  other
possible pentaquarks are also
discussed in the literature~\cite{Roca:2015dva,He:2015cea,He:2016pfa,Chen:2015loa,Yang:2011wz,Maiani:2015vwa,Liu:2015fea,Chen:2015moa,Meissner:2015mza,Lebed:2015tna,Chen:2016ryt}.

The hadronic molecular state picture is one of the most important interpretations to 
 explain LHCb pentaquarks as other resonance structures
which cannot be put into a conventional quark
model~\cite{Roca:2015dva,He:2015cea,He:2016pfa,Chen:2015loa,Yang:2011wz}.
Since the $P_c(4380)$ and $P_c(4450)$ are close to the $\Sigma_c^*
D$ and $\Sigma_c D^*$ thresholds,  it is natural to relate   two LHCb
pentaquarks  to the  $\Sigma_c^* D$ and $\Sigma_c D^*$ interactions.
In Ref.~\cite{He:2015cea},  a calculation in a quasipotential
Bethe-Salpeter approach suggested that a  bound state with quantum
number $J^P=3/2^-$ and a bound state with $5/2^+$ can be produced from
the  $\Sigma_c^* D$  and the $\Sigma_c D^*$ interactions,
respectively.  It is consistent with the experimental observation
about the hidden-charmed pentaquarks at LHCb. The $P_c(4450)$ is a
P-wave state in this picture, and an explicit study in
Ref.~\cite{He:2016pfa} showed that the S-wave state from the same interaction is
located around the $P_c(4380)$, which suggests that the $P_c(4380)$
may be a mixing state from two interactions.  Generally speaking, two
bound states produced from the $\Sigma_c^* D$ and $\Sigma_c D^*$
interactions can be related to the $P_c(4380)$ and $P_c(4450)$,
respectively.

It is interesting to go back to the light sector again. In fact, some
predictions about the hidden-charmed pentaquarks~\cite{Wu:2010jy,
Yuan:2012wz} were invoked by a  possible pentaquark component in the nucleon and its
resonance~\cite{Zou:2005xy}. The possible pentaquark composed of light
quarks has a longer history than that composed of  heavy quarks. The
hyperon resonance $\Lambda (1405)$ was explained as an $N\bar{K}$
bound state by many authors since the 1960s~\cite{Dalitz:1960du,Oller:2000fj, Kaiser:1995eg,Oset:1997it,Jido:2010ag,Hall:2014uca,Krippa:1998us,Mai:2014xna}.
In the chiral unitary approach, the interpretation of the
$\Lambda(1405)$ has been extended to other nucleon resonances, such as
$N(1535)$ and $N(1650)$~\cite{Inoue:2001ip,Mai:2012wy}. The nucleon
resonances are another
important issue of hadron physics, for example, the ``missing
resonances" problem. Until now, the nucleon resonances near 2 GeV
were
still unclear in both experiment and theory. Four $N(3/2^-)$ states,
$N(1520)$, $N(1700)$, $N(1875)$ and $N(2120)$, are listed in new
versions of the Review of Particle Physics (PDG) after the year
2012~\cite{Agashe:2014kda}.  The two-star state $N(2080)$ in previous
versions has been split into a three-star $N(1875)$ and a two-star
$N(2120)$ based on the evidence from BnGa
analysis~\cite{Anisovich:2011fc}. The interpretations about the
internal structure of  $N(1875)$ and $N(2120)$ are still  diverse in
the literature~\cite{Anisovich:2011fc,He:2012ud,He:2015yva}.

Many analyses have suggested that an $N(3/2^-)$ state with mass about 2.1
GeV is essential to explain experimental
results~\cite{Kohri:2009xe,Kim:2011rm,Xie:2010yk,He:2012ud}. Before
the year 2012, it was related  only to a state with spin parity $3/2^-$, listed in 
PDG with mass higher than 1.8 GeV, the $N(2080)$, and explained as the
third state  predicted in the constituent quark model~\cite{He:2012ud,He:2014gga}.    Recently, the CLAS Collaboration
at Jefferson National Accelerator Facility released their exclusive
photoproduction cross section for the $\Lambda$(1520) at energies
from near threshold upto a center-of-mass energy $W$ of 2.85 GeV in a 
large range of the $K$ production angle~\cite{Moriya:2013hwg}.  The
reanalyses about the new data in Refs.~\cite{He:2014gga,Xie:2013mua}
confirmed the previous conclusion that a nucleon resonance near 2.1
GeV, $N(2120)$, is essential to reproduce the experimental
data~\cite{Xie:2010yk,He:2012ud} [here and hereafter, we use $N(2120)$
to denote the nucleon resonance in the $K\Lambda(1520)$
photoproduction only]. An explicit calculation suggested that it can
be well explained as the third nucleon resonance state $[3/2^-]_3$ in the
constituent quark model~\cite{He:2012ud}.

The structure near 2.1 GeV can be tracked to an enhancement in the same
energy region in the $\phi$
photoproduction\cite{Mibe:2005er,Kiswandhi:2010ub,Kiswandhi:2011cq}
[we denote it as $N(2100)$ thereafter to avoid confusion with the
nucleon resonance $N(2120)$ in the $K\Lambda(1520)$ photoproduction and
the $N(2080)$ in the previous version of PDG]. A recent analysis about
the LEPS and CLAS data ~\cite{Chang:2010dg, Chang:2009yq, Qian:2010rr}
suggested that it has a mass of $2.08\pm0.04$ GeV and quantum number
of $J^P=3/2^-$~\cite{Kiswandhi:2016cav}. Since the two structures are
close to each other,  it is natural to think that  they have the same
origin. However, previous calculations in the constituent quark model
suggested that  the $N(2120)$ in the $K\Lambda(1520)$ channel  can be
well explained as the third state $[N3/2^-]_3$ predicted in the
constituent quark model~\cite{He:2012ud,He:2014gga}. The nucleon
resonance is composed of three nonstrange quarks in a conventional
quark model. It is difficult to produce in the $\phi$ photoproduction
because the $\phi$ meson is a particle with hidden strangeness and
produced without associated hyperons in this process, which leads to
serious suppression according to the OZI rule.   In fact, such
experiment is motivated by the idea to test the effect of the gluons
\cite{Seraydaryan:2013ija}. However, if  we assume that the $N(2100)$ is a
hidden-strangeness pentaquark instead of  a naive three-quark state,
the OZI suppression does not exist in either  production or
 decay. For example, in Ref.~\cite{Lebed:2015fpa}, the author
 suggested that the
enhancement in the $\phi$ photoproduction can be explained by 
production of  recoiling $su$ diquarks and $\bar{s}ud$
triqurks.

It is difficult to take the $N(1875)$   as a three-quark state in the
constituent quark model also. In our previous
works~\cite{He:2012ud,He:2014gga}, the $N(2120)$ in the
$K\Lambda(1520)$ photoproduction was assigned as a naive three-quark
state in the constituent quark model, so that there is no position to
settle the $N(1875)$, which is listed in  PDG as the third $N(3/2^-)$
nucleon resonance. Hence,  an interpretation was proposed that the
$N(1875)$ is from an interaction of a decuplet baryon $\Sigma(1385)$ and
an octet meson $K$, which is favored  by a calculation of binding
energy and decay pattern in a quasipotential Bethe-Salpeter approach
for the vertex~\cite{He:2015yva}. A study in the chiral unitary
approach also suggested a small peak near the $\Sigma^*K$
threshold~\cite{Sarkar:2004jh}.

Based on the  above analysis, it is difficult to put  either
$N(1875)$ or $N(2100)$ into the conventional quark model. If we
compare  $N(1875)$ and  $N(2100)$ with  LHCb pentaquarks $P_c(4380)$
and $P_c(4450)$, many similarities can be found. The two nucleon
resonances are close to the $\Sigma^* K$ and $\Sigma K^*$ thresholds
as LHCb pentaquarks to the $\Sigma_c^* D$ and $\Sigma_c D^*$
thresholds, and the $N(2100)$ was observed in the $\phi N$ channel as
LHCb pentaquarks in the $J/\psi N$ channel [the $N(1875)$ is below the
$\phi N$ threshold, so its decay is forbidden in this channel].
Hence, it is interesting to study if the $N(1875)$  and the $N(2100)$
are the strangeness partners of the LHCb pentaquarks in the hadronic
molecular picture. In this work, we investigate the possibility of
interpreting two nucleon resonances, the $N(1875)$ and the $N(2100)$, as
hadronic molecular states from the $\Sigma^*K$ and $\Sigma K^*$
interactions, respectively.

This paper is organized as follows.  After the introduction, we will
present  effective Lagrangians and  corresponding coupling constants
which are determined by the SU(3) symmetry.  In Sec.~\ref{Sec:
interaction}, the $\Sigma^*K$ and $\Sigma K^*$ interactions will be
given explicitly and the  quasipotential Bethe-Salpeter approach will
be introduced briefly, and then  adopted to study the interactions.
The coupled-channel effect from the coupling of the $\Sigma^*K$ and
$\Sigma K^*$ channels is also considered in our calculation. In
Sec.~\ref{Sec: results},  bound states are searched for and compared
with  experimentally observed nucleon resonances. Finally, the paper
ends with a discussion and conclusion.

\section{Effective Lagrangian}\label{Sec. Lagrangian}

To describe the $\Sigma^*K$ and $\Sigma K^*$ interactions in the
one-boson-exchange model, we should introduce  effective
Lagrangians for the vertices, whose coupling constants will be
determined with the help of the SU(3) symmetry following the  method
of de Swart~\cite{deSwart:1963pdg}.

For the $\Sigma^* K$ interaction we will consider exchanges of 
vector $\rho$,  $\omega$, and $\phi$ mesons.  The pseudoscalar-meson
exchanges are forbidden because the $K$ meson is also a pseudoscalar meson.
Different from the  charmed sector in which the exchange of 
hidden-charmed $J/\psi$ meson is not included~\cite{He:2015cea}, here
we include the hidden-strangeness $\phi$ meson because its mass is close to
other vector mesons, $\rho$ and $\omega$ mesons. We need the
Lagrangians for the vertices of  strange $K$ meson and vector mesons
as
\begin{eqnarray}
	{\cal L}_{KK\rho}&=&-ig_{KK\rho} ~K
	{\bm \rho}^\mu\cdot{\bm \tau} \partial_\mu K,\\
	{\cal L}_{KK\omega}&=&-ig_{KK\omega} ~K
	\omega^\mu \partial_\mu K,\\
	{\cal L}_{KK\phi}&=&-ig_{KK\phi} ~K
	\phi^\mu \partial_\mu K,
\end{eqnarray}
where the coupling constants are obtained by the SU(3) symmetry as
$g_{KK\rho}=g_{\rho\pi\pi}/2$, $g_{KK\omega}=g_{\rho\pi\pi}/2$, and
$g_{KK\phi}=g_{\rho\pi\pi}/\sqrt{2}$.  Here, we adopt a standard
$\alpha_{PPV}=F/(D+F)=1$ for vertices involving pseudoscalar
mesons~\cite{Janssen:1996kx,Ronchen:2012eg}. The  value of the
$g_{\rho\pi\pi}$ is determined as 6.1994 by the EBAC
group~\cite{Matsuyama:2006rp} and as 6.04 in Ref.~\cite{Ronchen:2012eg}.  We adopt a value of 6.1 in this work.

Besides the Lagrangians for the vertices of  strange $K$ meson and
vector mesons, the Lagrangians for the vertices of  strange $\Sigma^*$
baryon and vector mesons are also required and read
\begin{eqnarray}
	{\cal
	L}_{\Sigma^*\Sigma^*\rho}&=&-g_{\Sigma^*\Sigma^*\rho}~
	\bar{\Sigma}^{*\mu}[\gamma^\nu-\frac{\kappa_{\Sigma^*\Sigma^*\rho}}{2m_{\Sigma^*}}
	\sigma^{\nu\rho}\partial_\rho]{\bm\rho}_\nu\cdot{\bm T} \Sigma^*_\mu,\\
	{\cal
	L}_{\Sigma^*\Sigma^*\omega}&=&-g_{\Sigma^*\Sigma^*\omega}~
	\bar{\Sigma}^{*\mu}[\gamma^\nu-\frac{\kappa_{\Sigma^*\Sigma^*\omega}}{2m_{\Sigma^*}}
	\sigma^{\nu\rho}\partial_\rho]\omega_\nu \Sigma^*_\mu,\\
	{\cal
	L}_{\Sigma^*\Sigma^*\phi}&=&-g_{\Sigma^*\Sigma^*\phi}~
	\bar{\Sigma}^{*\mu}[\gamma^\nu-\frac{\kappa_{\Sigma^*\Sigma^*\phi}}{2m_{\Sigma^*}}
	\sigma^{\nu\rho}\partial_\rho]\phi_\nu \Sigma^*_\mu,
\end{eqnarray}
The relations between the  above coupling constants and the $g_{\Delta\Delta\rho}$ are determined by the SU(3) symmetry as $g_{\Sigma^*\Sigma^*\rho}=g_{\Delta\Delta\rho}$, $g_{\Sigma^*\Sigma^*\omega}=-g_{\Delta\Delta\rho}$,
and $g_{\Sigma^*\Sigma^*\phi}=g_{\Delta\Delta\rho}/\sqrt{2}$.  The matrix $T$  is defined as in Ref.~\cite{Matsuyama:2006rp}, and the values of  $g_{\Delta\Delta\rho}$ and $\kappa_{\Delta\Delta\rho}$ are chosen as $6.1994$ and $6.1$, respectively, as in the same reference.

For the $\Sigma K^*$ interaction,  exchanges of  vector mesons and of
pseudoscalar mesons should be included.  The Lagrangians for the
vertices
of strange vector $K^*$ meson and vector mesons are\small
\begin{align}
	{\cal L}_{K^*K^*\rho}&=i\frac{g_{K^*K^*\rho}}{2}( K^{*\mu}{\bm \rho}_{\mu\nu}K^{*\nu}+K^{*\mu\nu}{\bm \rho}_{\mu}K^{*\nu}+K^{*\mu}{\bm \rho}_{\nu}K^{*\nu\mu})\cdot{\bm \tau},\nonumber\\
	{\cal L}_{K^*K^*\omega}&=i\frac{g_{K^*K^*\omega}}{2} (K^{*\mu}{\omega}_{\mu\nu}K^{*\nu}+K^{*\mu\nu}{\omega}_{\mu}K^{*\nu}+K^{*\mu}{\omega}_{\nu}K^{*\nu\mu}),\nonumber\\
	{\cal L}_{K^*K^*\phi}&=i\frac{g_{K^*K^*\phi}}{2} (K^{*\mu}{\phi}_{\mu\nu}K^{*\nu}+K^{*\mu\nu}{\phi}_{\mu}K^{*\nu}+K^{*\mu}{\phi}_{\nu}K^{*\nu\mu}),
\end{align}\normalsize
where
$g_{K^*K^*\rho}=K^*K^*\omega=g_{K^*K^*\phi}/\sqrt{2}=g_{\rho\rho\rho}/2$
under the SU(3) symmetry,  and $g_{\rho\rho\rho}=g_{\rho\pi\pi}$~\cite{Bando:1987br,Janssen:1994uf}.  
The Lagrangians for the vertices of  strange $K^*$ meson and
pseudoscalar mesons are
\begin{eqnarray}
	{\cal
	L}_{K^*K^*\pi}&=&g_{K^*K^*\pi}~\epsilon^{\mu\nu\alpha\beta}\partial_\mu
	K^{*}_\nu\partial_\alpha{\bm \pi}\cdot{\bm \tau} K^{*}_\beta,\nonumber\\
	{\cal
	L}_{K^*K^*\eta}&=&g_{K^*K^*\eta}~\epsilon^{\mu\nu\sigma\tau}\partial_\mu
	K^{*}_\nu\partial_\alpha\eta K^{*}_\beta,
\end{eqnarray}
where  $g_{K^*K^*\pi}=g_{K^*K^*\eta}/\sqrt{3}=g_{\omega\rho\pi}/2$ under SU(3) symmetry with $g_{\omega\pi\rho}$ being 11.2~\cite{Matsuyama:2006rp}.

For the vertices of  strange $\Sigma$ baryon and vector mesons, the Lagrangians  read
\begin{eqnarray}
	{\cal
	L}_{\Sigma\Sigma\rho}&=&-g_{\Sigma\Sigma\rho}
	\bar{\Sigma}[\gamma^\nu-\frac{\kappa_{\Sigma\Sigma\rho}}{2m_{\Sigma}}
	\sigma^{\nu\rho}\partial_\rho]{\bm\rho}_\nu\cdot{\bm T} \Sigma,\nonumber\\
	{\cal
	L}_{\Sigma\Sigma\omega}&=&-g_{\Sigma\Sigma\omega}
	\bar{\Sigma}[\gamma^\nu-\frac{\kappa_{\Sigma\Sigma\omega}}{2m_{\Sigma}}
	\sigma^{\nu\rho}\partial_\rho]{\omega}_\nu \Sigma,\nonumber\\
	{\cal
	L}_{\Sigma\Sigma\phi}&=&-g_{\Sigma\Sigma\phi}
	\bar{\Sigma}[\gamma^\nu-\frac{\kappa_{\Sigma\Sigma\phi}}{2m_{\Sigma}}
	\sigma^{\nu\rho}\partial_\rho]{\phi}_\nu \Sigma,
\end{eqnarray}
where  coupling constants can be related to $g_{NN\rho}$ under the
SU(3) symmetry as  $g_{\Sigma\Sigma\rho}=2\alpha g_{NN\rho}$,
$g_{\Sigma\Sigma\omega}=2\alpha g_{NN\rho}$,   and
$g_{\Sigma\Sigma\phi}=-\sqrt{2}(2\alpha-1) g_{NN\rho}$.   The
$g_{NN\rho}$ is chosen as $g_{\rho\pi\pi}/2$ as in
Refs.~\cite{Matsuyama:2006rp, Ronchen:2012eg}, and we adopt a value of
$\alpha_{BBV}$ 1.15 as determined with coupled-channel reactions
$\pi N \to \pi N, \eta N, K\Lambda, K\Sigma$~\cite{Ronchen:2012eg}.
Under SU(3) symmetry, the $\kappa$ can be obtained with  relations
$f_{\Sigma\Sigma\rho}=(f_{NN\omega}+f_{NN\rho})/2$,
$f_{\Sigma\Sigma\omega}=(f_{NN\omega}+f_{NN\rho})/2$,
and $f_{\Sigma\Sigma\phi}=(-f_{NN\omega}+f_{NN\rho})/\sqrt{2}$,  where
$f_{BBV}$ is defined as $f_{BBV}=g_{BBV}\kappa_{BBV}$, and
$\kappa_\rho=6.1$ and $f_{NN\omega}=0$~\cite{Ronchen:2012eg}.

The Lagrangians for the vertices of  strange $\Sigma$ baryon and
pseudoscalar mesons are of the forms
\begin{eqnarray}
	{\cal
	L}_{\Sigma\Sigma\pi}&=&-\frac{f_{\Sigma\Sigma\pi}}{m_\pi}
	\bar{\Sigma}\gamma^5\gamma^\mu \partial_\mu {\bm \pi}\cdot{\bm T} \Sigma,\nonumber\\
	{\cal
	L}_{\Sigma\Sigma\eta}&=&-\frac{f_{\Sigma\Sigma\eta}}{m_\pi}
	\bar{\Sigma}\gamma^5\gamma_\mu \partial_\mu \eta \Sigma,
\end{eqnarray}
where under the SU(3) symmetry $f_{\Sigma\Sigma\pi}=2\alpha f_{NN\pi}$ and  $f_{\Sigma\Sigma\eta}=\frac{2}{\sqrt{3}}(1-\alpha)f_{NN\pi}$, and $\alpha=0.4$ and $f_{NN\pi}=1$~\cite{Ronchen:2012eg} .

In this  work, we will introduce the coupled-channel effect  from the coupling of  the $\Sigma^* K$ and the $\Sigma K^*$ channels considered in the above. The pseudoscalar-  and vector-meson exchanges will be introduced to describe the $\Sigma K^*-\Sigma^* K$ interaction.
Hence, we need the Lagrangians for the vertices of the strange mesons
and
vector mesons, 
\begin{eqnarray}
	{\cal L}_{K^*K\rho}&=&g_{K^*K\rho}\epsilon^{\mu\nu\sigma\tau}\partial_\mu K^{*}_{\nu}
	{\bm \rho}\cdot{\bm \tau} \partial_\sigma K,\nonumber\\
	{\cal L}_{K^*K\omega}&=&g_{K^*K\omega}\epsilon^{\mu\nu\sigma\tau}\partial_\mu K^{*}_{\nu}
	\omega \partial_\sigma K,\nonumber\\
	{\cal L}_{K^*K\phi}&=&g_{K^*K\phi}\epsilon^{\mu\nu\sigma\tau}\partial_\mu K^{*}_{\nu}
	\phi \partial_\sigma K,
\end{eqnarray}
where the relations
$g_{K^*K\rho}=g_{K^*K\omega}=g_{K^*K\phi}/[\sqrt{2}(2\alpha-1)]=g_{\omega\rho\pi}/(2\alpha)$
can be obtained with the SU(3) symmetry with
$\alpha_{VVV}=1$~\cite{Ronchen:2012eg}. The Lagrangians for the
vertices of the strange mesons and pseudoscalar mesons read
\begin{eqnarray}
	{\cal L}_{K^*K\pi}&=&-ig_{K^*K\pi}K^{*\mu}
	({\bm \pi}\partial_\mu-\partial_\mu{\bm \pi})\cdot{\bm \tau} K,\nonumber\\
	{\cal L}_{K^*K\eta}&=&-ig_{K^*K\pi}K^{*\mu}
	({ \eta}\partial_\mu-\partial_\mu{\eta}) K,
\end{eqnarray}
with $g_{K^*K\pi}=- g_{\rho\pi\pi}/2$ and $g_{K^* K\eta}=-\sqrt{3}g_{\rho\pi\pi}/2$~\cite{Ronchen:2012eg} .

The Lagrangians for the vertices of the strange baryons and vector mesons  read
\begin{eqnarray}
	{\cal
	L}_{\Sigma^*\Sigma\rho}&=&-i\frac{f_{\Sigma^*\Sigma\rho}}{m_\rho}
	\bar{\Sigma}^{*\mu}\gamma^5\gamma^\nu[\partial_\mu{\bm\rho}_\nu-\partial_\nu
	{\bm \rho}_\mu]\cdot{\bm T} \Sigma,\nonumber\\
	{\cal
	L}_{\Sigma^*\Sigma\omega}&=&-i\frac{f_{\Sigma^*\Sigma\omega}}{m_\rho}
	\bar{\Sigma}^{*\mu}\gamma^5\gamma^\nu[\partial_\mu{\omega}_\nu-\partial_\nu
	{ \omega}_\mu] \Sigma,\nonumber\\
	{\cal
	L}_{\Sigma^*\Sigma\phi}&=&-i\frac{f_{\Sigma^*\Sigma\phi}}{m_\rho}
	\bar{\Sigma}^{*\mu}\gamma^5\gamma^\nu[\partial_\mu{\phi}_\nu-\partial_\nu
	{ \phi}_\mu] \Sigma,
\end{eqnarray}
where $f_{\Sigma^*\Sigma\rho}=-f_{\Delta N \rho}/\sqrt{6}$,
$f_{\Sigma^*\Sigma\omega}=-f_{\Delta N \rho}/\sqrt{2}$,  and
$f_{\Sigma^*\Sigma\phi}=-\sqrt{2}f_{\Delta N \rho}$ with $f_{\Delta N\rho}=-6.08$~\cite{Matsuyama:2006rp}.

The Lagrangians for the vertices of the strange  baryons and pseudoscalar mesons  read
\begin{eqnarray}
	{\cal
	L}_{\Sigma^*\Sigma\pi}&=&\frac{f_{\Sigma^*\Sigma\pi}}{m_\pi}
	\bar{\Sigma}^{\mu}\partial_\mu {\bm \pi}\cdot{\bm T} \Sigma,\nonumber\\
	{\cal
	L}_{\Sigma^*\Sigma\eta}&=&\frac{f_{\Sigma^*\Sigma\eta}}{m_\pi}
	\bar{\Sigma}^\mu \partial_\mu \eta \Sigma,
\end{eqnarray}
where $f_{\Sigma^*\Sigma\pi}=-f_{\Sigma^*\Lambda\pi}/\sqrt{3}$ 
and  $f_{\Sigma^*\Sigma\eta}=-f_{\Sigma^*\Lambda\pi}$ with $f_{\Sigma^*\Lambda\pi}=1.27$~\cite{He:2015yva, Gao:2010ve}.

\section{$\Sigma^*K$ and $\Sigma K^*$ interactions}\label{Sec: interaction}

With the Lagrangians above, the potential kernel ${\cal V}$ of the
interactions can be obtained with the help of the standard  Feynman rule,
which includes dynamical information of the interactions, and will be
used to study possible bound states produced from the interactions. 

The potential for the 
$\Sigma^*K$ interaction by  exchanges of  vector $\mathbb{V}$  mesons
is written as
\begin{align}
	i{\cal V}_{\mathbb{V}}&=
f_I\frac{g_{KK\mathbb{V}}g_{\Sigma^*\Sigma^*\mathbb{V}}}{q^2-m_\mathbb{V}^2}
\bar{u}^\mu
\left\{-\rlap{$\slash$} k_1+\frac{q\cdot k_1
\rlap{$\slash$} q}{m_\mathbb{V}^2}-\frac{\kappa_{\Sigma^*\Sigma^*\mathbb{V}}}{4m_{\Sigma^*}}
[\rlap{$\slash$}k_1,\rlap{$\slash$}q]\right\}u_\mu,
\end{align}
where $q=k'_1-k_1$, with $k_1$ and $k'_1$ being the momenta of initial
and final $K$ mesons, respectively, and $u^\mu$ is the  Rarita-Schwinger
vector-spinor for the strange baryon $\Sigma^*$. For the exchanges of the isovector $\rho$ and $\pi$ mesons, the isospin factor $f_I=-2$ and 1 for
isospin $1/2$ and $3/2$, respectively. For other exchanges, the isospin factor $f_I=1$.

The potential kernel ${\cal V}$ for the $\Sigma K^*$ interaction
by vector $\mathbb{V}$ and pseudoscalar  $\mathbb{P}$ exchanges can be written as
\begin{align}
	i{\cal V}_{\mathbb{V}}&=f_I\frac{g_{K^*K^*\mathbb{V}}g_{\Sigma\Sigma\mathbb{V}}}{2}
	[\epsilon^{\dag}\cdot q
	\epsilon^{\nu}+(k_1+k'_1)^\nu~\epsilon^{\dag}\cdot
	\epsilon\nonumber\\
	&-\epsilon^{\dag\nu} \epsilon\cdot q-k_1\cdot \epsilon
	\epsilon^{\dag^\nu}-k'_1\cdot \epsilon^{\dag}\epsilon^{\nu}]
	\nonumber\\&\cdot
\frac{g_{\nu\nu'}-q_\nu
q_{\nu'}/m_\mathbb{V}^2}{q^2-m_\mathbb{V}^2}
~\bar{u}	(\gamma^{\nu'}-i\frac{\kappa_{\Sigma\Sigma\mathbb{V}}}{2m_\Sigma}
\sigma^{\nu'\rho}q_\rho)u\nonumber\\
i{\cal V}_{\mathbb{P}}&=f_I\frac{g_{K^*K^*\mathbb{P}} f_{\Sigma\Sigma\mathbb{P}}}{m_\pi(q^2-m_\mathbb{P}^2)}
~\epsilon^{\mu\nu\alpha\beta}k'_{1\mu} \epsilon^{\dag}_\nu
k_{1\alpha} \epsilon_{\beta}
	~\bar{u}\gamma_5\rlap{$\slash$} q  u,
\end{align}
where  $\epsilon$ and $u$ are the polarized vector for the strange $K^*$ meson and
the spinor for the  strange  $\Sigma$ baryon, respectively. 

The potential kernel ${\cal V}$ for the coupling of the $\Sigma^* K$ and $\Sigma K^*$ channels can be written as
\begin{eqnarray}
	i{\cal
	V}_{\mathbb{V}}
&=&f_Ig_{K^*K\mathbb{V}}\frac{f_{\Sigma^*\Sigma\mathbb{V}}}{m_\mathbb{V}}\epsilon^{\mu\nu\sigma\tau}k_{1\mu} \epsilon_{\nu}q_\sigma\nonumber\\
&\cdot&
	[\bar{u}^{\rho}\gamma^{\tau}q_\rho-\bar{\Sigma}^{\tau}\gamma^{\rho}q_\rho]\gamma_5 u\frac{1}{q^2-m^2_\mathbb{V}},\nonumber\\
i{\cal V}_{\mathbb{P}}
&=&if_Ig_{K^*K\pi}\frac{f_{\Sigma^*\Sigma\mathbb{P}}}{m_\mathbb{P}}\epsilon^{\mu}(k'_1+q)_\mu\bar{u}^{\nu}q_\nu u\frac{1}{q^2-m_\mathbb{P}^2}.
\end{eqnarray}

Now we have the potential kernels of the $\Sigma^* K$ interaction, the
$\Sigma K^*$ interaction, and their coupling with the parameters
fixed by the SU(3) symmetry.  To obtain the interaction amplitude, we
introduce the widely adopted Bethe-Salpeter equation. To solve the
Bethe-Salpeter equation, a spectator quasipotential approximation will
be adopted by putting one of the two particles
on shell~\cite{Gross:2008ps,Gross:2010qm,He:2014nya,He:2013oma}. As
discussed in Ref.~\cite{Gross:1999pd}, the heavier particle, here the
strange baryon, should be put on shell in this work because one-boson
exchange  is adopted.  A simple test of different choices of the
on-shell particle will also be made in this work. We would like to
remind the reader that the covariance and  unitary are still satisfied in this
approach.  The method was explained explicitly in the appendixes of
Ref.~\cite{He:2015mja}, and it has been applied to study the LHCb pentaquarks and
other exotic
states~\cite{He:2015cea,He:2014nxa,He:2016pfa,He:2015cca}. 

The
molecular state produced from the $\Sigma^* K$ and $\Sigma K^*$
interaction corresponds to a pole of the scattering amplitude ${\cal
M}$. The  quasipotential Bethe-Salpeter equation for partial-wave
amplitude with fixed spin-parity $J^P$ reads
~\cite{He:2015mja,He:2015cea}
\begin{eqnarray}
i{\cal M}^{J^P}_{\lambda'\lambda}({\rm p}',{\rm p})
&=&i{\cal V}^{J^P}_{\lambda',\lambda}({\rm p}',{\rm
p})+\sum_{\lambda''\ge0}\int\frac{{\rm
p}''^2d{\rm p}''}{(2\pi)^3}\nonumber\\
&\cdot&
i{\cal V}^{J^P}_{\lambda'\lambda''}({\rm p}',{\rm p}'')
G_0({\rm p}'')i{\cal M}^{J^P}_{\lambda''\lambda}({\rm p}'',{\rm
p}),\quad\quad \label{Eq: BS_PWA}
\end{eqnarray}
where 
with the potential kernel ${\cal V}_{\lambda'\lambda}$ obtained in the previous section,  the partial wave potential with fixed spin-parity $J^P$ can be calculated   as
\begin{eqnarray}
i{\cal V}_{\lambda'\lambda}^{J^P}({\rm p}',{\rm p})
&=&2\pi\int d\cos\theta
~[d^{J}_{\lambda\lambda'}(\theta)
i{\cal V}_{\lambda'\lambda}({\bm p}',{\bm p})\nonumber\\
&+&\eta d^{J}_{-\lambda\lambda'}(\theta)
i{\cal V}_{\lambda'-\lambda}({\bm p}',{\bm p})],
\end{eqnarray}
where without loss of  generality  the initial and final relative
momenta can be chosen as ${\bm p}=(0,0,{\rm p})$  and ${\bm p}'=({\rm
p}'\sin\theta,0,{\rm p}'\cos\theta)$ with a definition ${\rm
p}^{(')}=|{\bm p}^{(')}|$, and $d^J_{\lambda\lambda'}(\theta)$ is the
Wigner d-matrix. It is easy to extend above the one-channel equation
to the  coupled-channel case as in Refs.~\cite{He:2015cca,He:2016pfa}.

In this work we will introduce an  exponential regularization by
adding a form factor in the propagator as 
 \begin{eqnarray}
 G_0({\rm p})\to G_0({\rm
 p})\left[e^{-(k_1^2-m_1^2)^2/\Lambda^4}\right]^2,\label{Eq:
 FFG} 
 \end{eqnarray} 
where $k_1$ and $m_1$ are the momentum and mass of the strange meson,
respectively.
The interested reader is referred to Ref.~\cite{He:2015mja} for
further information about the regularization. Besides the exponential
regularization, we also introduce a direct cutoff in the one-channel calculation to check the
validity of the exponential regularization; that is, we cut off the
momentum ${\rm p}''$ at a value ${\rm p}^{max}$, which corresponds to the  cutoff regularization in the chiral unitary approach~\cite{Oller:1998hw}. Because such
treatments   guarantee the convergence of the integration, we do
not introduce the form factor for the exchanged meson, which is
redundant, and its effect can be absorbed into the small variation of the cutoffs as discussed
in Ref.~\cite{Lu:2016nlp}.

\section{Numerical results}\label{Sec: results}

With the above preparation, the bound states from the $\Sigma^*
K-\Sigma K^*$ interactions can be studied by searching for the pole of
the scattering amplitude. In the current work, the coupling constants
in the Lagrangians are determined by the SU(3) symmetry, so the only
free parameter is the cutoff $\Lambda$ for  exponential regularization
or ${\rm p}^{max}$ for  cutoff regularization. The cutoff $\Lambda$ should
note be  far from 1 GeV, and  ${\rm p}^{max}$ should be near 1 GeV as in
the chiral unitary
approach~\cite{Jido:2010ag,Inoue:2001ip,Oller:1998hw}. We  allow the
cutoffs to deviate a little as in the chiral unitary approach to
absorb the small effects, which is not included in our formalism. We
will investigate all quantum numbers with $J=1/2$, $3/2$, and $5/3$  in
a range of $\Lambda$ from 0.8 to 2.5 GeV. The corresponding ${\rm
p}^{max}$ will also be given.  

\subsection{$\Sigma^*K$ interaction}

We will present first the results for the one-channel calculations for
$\Sigma^*K$ interaction and $\Sigma K^*$ interaction in this and
the following subsection, respectively. In Table~\ref{Tab: bound state},
the bound states produced from the $\Sigma^*K$ interaction are listed.
For the one-channel interaction, the pole for the bound state is at
the real axis.

\renewcommand\tabcolsep{0.216cm}
\renewcommand{\arraystretch}{1.7}
\begin{table}[h!]
\begin{center}
\caption{The bound states from the $\Sigma^*K$ interaction with the variation of the cutoffs $\Lambda$ or ${\rm
p}^{max}$ . The cutoff $\Lambda$, cutoff ${\rm p}^{max}$, and  energy $W$ are in units of GeV, GeV, and MeV, respectively. \label{Tab: bound state}
\label{diagrams}}
	\begin{tabular}{c|ccc||c|cccccc}\bottomrule[1.5pt]
$I(J^{P})$   &  $\Lambda$ &${\rm p}^{max}$ &  $W$&$I(J^{P})$   &  $\Lambda$ &${\rm p}^{max}$ &  $W$  \\\hline
$\half(\thalf^-)$ 
& 1.5  & 1.10 & 1880  &$\thalf(\half^+)$ &1.60& 1.238 &1878 \\
& 1.6  & 1.15 & 1879  && 1.61& 1.249 &1870 \\
& 1.7  & 1.25 & 1874  &&  1.62& 1.259 &1862 \\
& 1.8  & 1.34 & 1862  && 1.63& 1.269 &1854\\
& 1.9  & 1.56& 1832  &&  1.64&1.280 &1844\\
\toprule[1.5pt]
\end{tabular}
\end{center}

\end{table}

The numerical calculation suggests that in the isospin $1/2$ sector,
only one bound state with quantum number $I(J^{P}) = 1/2(3/2^-)$ can
be reproduced from the $\Sigma K^*$ interaction. It is consistent with
our previous study with a Bethe-Salpeter equation for the vertex where
we also found only one state in this sector~\cite{He:2015yva}. Obviously, this
bound state corresponds  to the nucleon resonance $N(1875)$
listed in PDG~\cite{Agashe:2014kda}. In the isospin $3/2$ sector, we
also find only one bound state with quantum number $3/2(1/2^+)$. If we
adopt the cutoff ${\rm p}^{max}$,  analogous results can be obtained,
and the result suggests that a cutoff $\Lambda$ about 1.6 GeV
corresponds to a cutoff ${\rm p}^{max}$ about 1.2 GeV. The values of
both cutoffs for the two bound states are reasonable.  

Since the one-boson-exchange model is adopted to describe the
interaction, the charge-conjugation invariance requires that the
heavier particle, here the strange baryon, should be put
on shell~\cite{Gross:1999pd}. However, it is interesting to present the
results with the lighter particle onshell to test the reliability of
the quasipotential method.  Hence, a calculation with the strange
meson on shell  is made here to compare the results with two choices as
listed in Table \ref{Tab: bound state test}. We vary the cutoffs to
obtain the same  energy as the case with the baryon on shell. It is found
that  almost the same result can be obtained after making a variation
of the cutoff $\Lambda$. In  other words, the choice of the on-shell
particle does not affect the conclusion.

\renewcommand\tabcolsep{0.199cm}
\renewcommand{\arraystretch}{1.7}
\begin{table}[h!]
\begin{center}
\caption{The bound states from the $\Sigma^*K$ interaction at different cutoffs $\Lambda$ or ${\rm
p}^{max}$ with the strange meson on shell. The cutoff $\Lambda$, cutoff ${\rm p}^{max}$, and  energy $W$ are in units of GeV, GeV, and MeV, respectively. \label{Tab: bound state test}
\label{diagrams}}
	\begin{tabular}{c|ccc||c|cccccccc}\bottomrule[1.5pt]
		$I(J^{P})$   &  $\Lambda$  &${\rm p}^{max}$ &  $W$&$I(J^{P})$   &  $\Lambda$  &${\rm p}^{max}$ &  $W$  \\\hline
$\half(\thalf^-)$ 
&1.40  &0.70& 1880  &$\thalf(\half^+)$ &1.58&0.925&1878 \\
 &1.44  &0.75& 1879  &&1.59 &0.931&1870 \\
&1.49  &0.81& 1874  &&1.60 &0.935&1862 \\
 &1.53 &0.93& 1862  &&1.61 &0.938&1854\\
&1.60 &1.06& 1832  &&1.62 &0.943&1844\\
\toprule[1.5pt]
\end{tabular}
\end{center}

\end{table}

When comparing the results of two bound states carefully, one can find
that the binding energy for the $3/2(1/2^+)$ bound state changes much
faster than the $1/2(3/2^-)$ bound state with a variation of the
cutoff. For the $3/2(1/2^+)$ state, a variation of $\Lambda$ of about 0.02
GeV will lead to an increase of binding energy of  about 20 MeV while
a variation of $\Lambda$ of about 0.3 GeV is needed to lead to such an
increase for the $1/2(3/2^-)$ state.   The physical cutoff is a fixed
value for an interaction channel (though we do not know the explicit
value),  and if the bound state is far from the corresponding
threshold, its effect will become smaller and unreliable in the
hadronic molecular state picture.   Hence, the possibility of the
existence of the $3/2(1/2^+)$ state   will be much smaller than that
of the $1/2(3/2^-)$ state.

\subsection{$\Sigma K^*$ interaction}
The bound states from the $\Sigma K^*$ interaction with variation of the cutoff are listed in
Table~\ref{Tab: bound state2}. 
\renewcommand\tabcolsep{0.189cm}
\renewcommand{\arraystretch}{1.7}
\begin{table}[h!]
\begin{center}
\caption{The bound states from the $\Sigma K^*$ interaction at different cutoffs $\Lambda$ or ${\rm
p}^{max}$. 
The cutoff $\Lambda$, cutoff ${\rm p}^{max}$, and  energy $W$ are in units of GeV, GeV, and MeV, respectively. \label{Tab:
bound state2}
\label{diagrams}}
	\begin{tabular}{c|ccc||c|cccccc}\bottomrule[1.5pt]
$I(J^{P})$   &  $\Lambda$ &${\rm p}^{max}$ &  $W$&$I(J^{P})$   &  $\Lambda$ &${\rm p}^{max}$ &  $W$  \\\hline
$\half(\thalf^-)$ 
& 0.8  & 0.46& 2086  &$\half(\fhalf^+)$  & 1.17  & 0.755& 2086      \\  
& 0.9  & 0.50 & 2085                 &   & 1.19  & 0.765 & 2082     \\  
& 1.0  & 0.57 & 2081                 &  & 1.21 & 0.781 & 2076      \\   
& 1.1  & 0.62 & 2076                 &   & 1.23 & 0.796 & 2069      \\  
& 1.2  & 0.69 & 2068                 &   & 1.25 & 0.808& 2060   \\\hline
$\half(\thalf^+)$ 
& 1.25  & 0.831& 2086 &$\thalf(\fhalf^+)$ & 1.31  & 0.910& 2087   \\     
& 1.26 & 0.839& 2084  &                   & 1.32 & 0.917 & 2084 \\       
& 1.27 & 0.845 & 2080 &                   & 1.33 & 0.935 & 2071  \\      
& 1.28 & 0.849 & 2076 &                   & 1.34 & 0.945 & 2061  \\      
& 1.29 & 0.854 & 2072 &                   & 1.29 & 0.953 & 2048 \\\hline 
 $\half(\fhalf^-)$ &1.43& 0.999 &2086  &$\thalf(\half^+)$ &0.93& 0.603 &2085  \\
&  1.44&1.007 &2080                    &                  & 0.94& 0.612 &2084  \\
&  1.45&1.015 &2071                    &                  &  0.95& 0.620 &2081\\
&  1.46&1.022 &2059                    &                  & 0.96& 0.625 &2078 \\
&  1.47&1.031 &2044                    &                  &  0.97&0.631 &2074 \\\hline
\toprule[1.5pt]
\end{tabular}
\end{center}

\end{table}

Different from the $\Sigma^* K$
interaction, there are four isospin 1/2 bound states and two isospin
3/2 bound states produced from the $\Sigma K^*$ interaction. The bound
state with quantum number $1/2(3/2^-)$ can be related to the $N(2100)$
in the $\phi$ photoproduction.  The result of such a state is also
stable with the variation of  cutoff $\Lambda$ or ${\rm
p}^{max}$. Other bound states will leave the $\Sigma K^*$ threshold
rapidly with an  increase of the cutoff, which suggests that the
possibility of their existence is smaller than the $1/2(3/2^-)$ bound state as discussed above.

\subsection{$\Sigma^*K-\Sigma K^*$ interaction}

The $\Sigma^*K$ and $\Sigma K^*$ channels can be connected with the
pseudoscalar and vector meson exchanges.  A coupled channel
calculation can  be made with the inclusion of the coupling of the
$\Sigma^*K$ and $\Sigma K^*$ channels.  Based on the analysis above,
only two bound states with $1/2(3/2^-)$ from the $\Sigma^* K$
interaction and $\Sigma K^*$ interaction are stable with the variation
of the cutoff, and they correspond to the $N(1875)$ and the $N(2120)$ in
the experiment, respectively.  Hence, here, we focus on the case of
$1/2(3/2^-)$ only. 

\begin{figure}[h!]
\begin{center}
\includegraphics[bb=75 90 400 260, clip, scale=0.8]{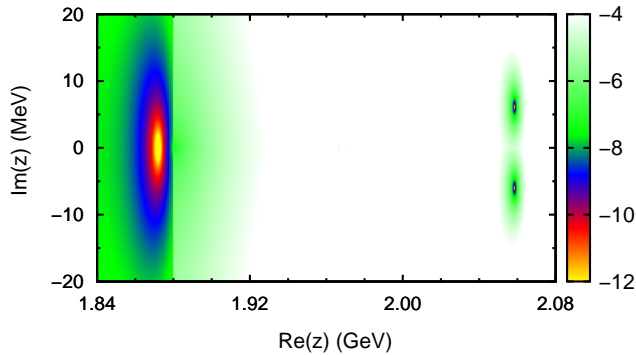}
\end{center}
\caption{$\log|1-V(z)G(z)|$ with the variation of  $z$ for the $\Sigma^*K-\Sigma K^*$ interaction. }
\label{Fig: pole}
\end{figure}

In the coupled-channel case,  two cutoffs, $\Lambda_{\Sigma^* K}$ and
$\Lambda_{\Sigma K^*}$,  will be involved for the $\Sigma^*K$ and
$\Sigma K^*$ channels, respectively. Since different baryons and
mesons are involved in two channels, it is unnatural to adopt the same
cutoff for two channels. Hence, in this work we will adopt different
cutoffs for different channels.  First,  we take the case with
$\Lambda_{\Sigma^* K}=1.7$ GeV and $\Lambda_{\Sigma K^*}=1.3$ GeV as
an example to illustrate the poles from the coupled-channel calculation
as shown in  Fig.~\ref{Fig: pole}. 

After inclusion of the coupled-channel effect, there are still two
poles produced near two thresholds, and it is obvious that the
higher and lower poles correspond to the $\Sigma K^*$ and $\Sigma^* K$
channels, respectively.  As expected, the higher pole from the $\Sigma
K^*$ channel will deviated form the real axis after inclusion of  the
coupling of two channels as shown in  Fig.~\ref{Fig: pole},  which
reflects the decay width from opening the decay of the $\Sigma^* K$
channel. The pole for the lower pole is still at  real axis because
no decay channel is open in the calculation.

In Table~\ref{Tab: bound state3}, the poles from the $\Sigma^*K-\Sigma
K^*$ interaction with the variations of two cutoffs are presented.
Empirically, the larger cutoff will lead to a larger coupling of two
channels as well as  the one-channel interaction. It is well reflected in
the results listed in Table~\ref{Tab: bound state3}. When we fix
$\Lambda_{\Sigma^* K}$ and increase $\Lambda_{\Sigma K^*}$, both poles
run farther away from the thresholds but from different origins.  The
upper pole, which is produced from the $\Sigma K^*$ interaction, runs
fast because the cutoff $\Lambda_{\Sigma K^*}$ affects the strength of the $\Sigma
K^*$ interaction directly. The lower pole runs  relatively slowly because
the cutoff $\Lambda_{\Sigma K^*}$ does not sffect the $\Sigma^* K$
interaction, which produces this pole. The origin of running of the
lower pole is the enhancement of the coupling of two channels. When we fix
$\Lambda_{\Sigma K^*}$ and increase $\Lambda_{\Sigma^* K}$, similar
phenomena can be found.  

\renewcommand\tabcolsep{0.13cm}
\renewcommand{\arraystretch}{1.7}
\begin{table}[h!]
\begin{center}
\caption{The poles from the $\Sigma^*K-\Sigma K^*$ interaction with
variations of the cutoffs $\Lambda_{\Sigma^* K}$ and $\Lambda_{\Sigma K^*}$. 
The cutoffs and  the positions of the poles are in units of GeV and MeV, respectively.  For each value of  $\Lambda_{\Sigma^* K}$, the higher and lower lines are for the higher and lower poles, respectively. \label{Tab:
bound state3}
\label{diagrams}}
	\begin{tabular}{c|lllll}\bottomrule[1.5pt]
\diagbox{$\Lambda_{\Sigma^* K}$}{ $\Lambda_{\Sigma K^*}$}  & 0.8 & 1.0
& 1.2 & 1.4 &1.6 \\\hline
 1.5  &  2086+$i$ &2081+$i$2 &2068+$i$4   &2046+$i$8 &1994+$i$12   \\  
      &  1880 &1879 &1879   &1878 &1878     \\  
 1.7  &  2086+$i$ &2081+$i$2 &2067+$i$4  &2046+$i$8 &1995+$i$11   \\  
      &  1874 &1874 &1873   &1871 &1869    \\  
 1.9  &  2086+$i$ &2081+$i$3 &2067+$i$5   &2046+$i$8 &1.995+$i$10    \\   
      &  1831 &1828 &1824   &1819 &1810     \\  
 2.1  &  2086+$i$ &2081+$i$3 &2067+$i$5   &2045+$i$7 &1993+$i$8    \\  
      &  $1786$ &$1779$  &$1768$    &$1759$  &$1740$      \\  
\toprule[1.5pt]
\end{tabular}
\end{center}

\end{table}

The width of the higher pole increases with the increase of cutoff
$\Lambda_{\Sigma K^*}$ at all fixed values of cutoff $\Lambda_{\Sigma^*
K}$, which is a mixing effect of the enhancement of the coupling of
two channels and fast running of the higher pole. When the cutoff
$\Lambda_{\Sigma^* K}$  increases at fixed $\Lambda_{\Sigma K^*}$, the width
decreases relatively slowly  because the upper pole is produced
from the $\Sigma K^*$ interaction, which is independent of the cutoff
$\Lambda_{\Sigma^* K}$.  Also, at larger $\Lambda_{\Sigma K^*}$, the width
becomes smaller with an increase of the cutoff $\Lambda_{\Sigma^*
K}$. It is reasonable because the width of the pole will increase
first and then decrease when the pole runs away from the threshold.  

Generally speaking, after inclusion of the
coupled-channel effect of the $\Sigma^*K$ and $\Sigma K^*$  channels,
the conclusion obtained with the one-channel calculation is 
unchanged; that is, two bound states are produced from the $\Sigma^*K$
and $\Sigma K^*$ channels, respectively.

\section{SUMMARY}

The previous studies suggested the $N(2120)$ in the $K\Lambda(1520)$
photoproduction is a three-quark state in the constituent quark
model~\cite{He:2012ud,He:2014gga}, which makes it difficult to put the
$N(1875)$ and $N(2100)$ in the $\phi$ photoproduction into the
constituent quark model.  Such difficulty and the closeness of the
$N(1875)$ and $N(2100)$  to the  $\Sigma^* K$ and  $\Sigma K^*$
thresholds invoke us to consider the possibility of interpreting  this
two nucleon resonances in the hadronic molecular state picture, which
is analogous to the LHCb pentaquarks. 

In the one-boson-exchange model, the interaction potentials are
obtained with the effective Lagrangians with the coupling constants
fixed by the SU(3) symmetry.  The bound states from the interactions
are studied through solving the quasipotential Bethe-Salpeter
equation. A bound state with quantum number $I(J^P)=1/2(3/2^-)$ from
the $\Sigma^* K$ interaction and a bound state with $1/2(3/2^-)$ from
the $\Sigma K^*$ interaction are produced with reasonable cutoffs.
These two bound states can be related to the $N(1875)$ and the
$N(2100)$, respectively.  The results for these two bound states are
also stable with the  variation of the cutoff.  Other bound states are
also produced from the two interactions. However, they leave the threshold
rapidly with an increase of the cutoff. Hence, though these bound
states can be produced in a narrow window of the cutoff, the
possibility of their existence in the real world is very small. 
The coupled-channel effect from the coupling of the $\Sigma^* K$ and
$\Sigma K^*$ channels is also discussed, and it is found that the
conclusion with the one-channel calculation is unchanged after
inclusion of the coupled-channel effect.

With such an assignment, three $3/2^-$ nucleon resonances, $N(1875)$,
$N(2100)$ in $\phi$ photoproduction, and $N(2120)$ in $K$
photoproduction with $\Lambda(1520)$, can be well understood.  The
two hadronic molecular states, $N(1875)$ and $N(2100)$, can be seen as
the strange partners of the LHCb pentaquarks.  The difference is that
in the hidden-strangeness sector, the two states are both in an S wave
while in the hidden-charm sector; we should put one state in a P wave to
reproduce the spin-parity suggested by the LHCb experiment.

\section*{Acknowledgments}

This project is partially supported by the National
Natural Science Foundation of China (Grant No.11675228) , and the Major State Basic Research
Development Program in China (Grant No. 2014CB845405).

\end{document}